\newcommand{\be}{\begin{equation}}\newcommand{\ee}{\end{equation}}
\newcommand{\bea}{\begin{eqnarray}}\newcommand{\eea}{\end{eqnarray}}
\newcommand{\nn}{\nonumber}
\documentstyle[12pt]{article}
\begin{document}
\renewcommand{\thefootnote}{\fnsymbol{footnote}}
\thispagestyle{empty}
%\vspace{3cm}
\begin{flushright}
JINR-E2-95-27\\
hep-th/9501100\\
Jan., 1995 \\
\end{flushright}
\vspace{2cm}
\begin{center}
THE FULL STRUCTURE OF QUANTUM $N=2$ SUPER-$W_3^{(2)}$ ALGEBRA \\
\vspace{1cm}
{\sc C. Ahn}\footnote{E-mail: ahn@thsun1.jinr.dubna.su},
{\sc S. Krivonos}\footnote{E-mail: krivonos@thsun1.jinr.
dubna.su} and
{\sc A. Sorin}\footnote{E-mail: sorin@thsun1.jinr.dubna.su} \\
\vspace{1cm}
{\it Bogoliubov  Theoretical Laboratory, \\
 JINR, 141 980, Dubna, Moscow region, Russia }\\
\vspace{3cm}
{\bf {\sc Abstract}}
\end{center}

We present the complete structure of the nonlinear
$N=2$ super extension of Polyakov-Bershadsky, $W_3^{(2)}$, algebra  with
the generic central charge, $c$, at the {\it quantum} level.
It contains extra two pairs of fermionic currents with integer spins 1 and 2,
besides the currents of $N=2$ superconformal and $W_3^{(2)}$ algebras.
For $c\rightarrow \infty$ limit, the algebra reduces to the classical
one, which has been studied previously.
The 'hybrid' field realization of this algebra is also discussed.
\vspace{1cm}
\vfill
\setcounter{page}0
\renewcommand{\thefootnote}{\arabic{footnote}}
\setcounter{footnote}0
\newpage

{\bf 1. Introduction}

Among the {\it nonlinear} bosonic algebras, there exist
very special algebras which contain the bosonic currents
with {\it noncanonical} half-integer spins \cite{p,b,r} contrary to other
algebras possessing the currents with canonical spins.
The Polyakov-Bershadsky, $W_3^{(2)}$, algebra \cite{p,b} is the simplest
nontrivial example of such algebras.
It is the bosonic analogue of the {\it linear}
$N=2$ superconformal algebra (SCA) \cite{x} and includes
two bosonic currents with noncanonical spins $3/2$ and two bosonic currents
with canonical spins $1, 2$.

Recently \cite{ks},
its $N=2$ supersymmetric extension has been constructed
at the {\it classical} level in the
sense that one should take into account
only single contractions between the composite
currents.
This algebra comprises,
besides
the currents of $W_3^{(2)} \propto \left\{ J_w, G^{+}, G^{-}, T_w \right\}$ and
$N=2$ superconformal $\propto \left\{ J_s, S, \bar{S}, T_s\right\} $
subalgebras with the same spins $\left(1,3/2,3/2,2 \right)$ respectively,
also additional four fermionic
currents $ \propto \left\{ S_1,{\bar S}_1,S_2,{\bar S}_2 \right\} $
with non-canonical integer spins $\left(1,1,2,2 \right)$:
the currents $ S,\bar{S}$ are fermionic, their counterparts $G^{+},G^{-}$
are bosonic. There is no  intersection of these
subalgebras at the embedding in this extended algebra, where
all the currents with integer spins can be obtained as
the right hand side of operator product expansions ( OPEs )
between those with half-integer spins.

In this paper, we  present  the $N=2$ supersymmetric extension of
the $W_3^{(2)}$ algebra at the {\it quantum} level, by taking into
account
Jacobi identities to all orders in contractions between the composite
currents, and construct explicitly
its 'hybrid' field realization on six bosonic and six
fermionic fields.

{\bf 2. The quantum $N=2$ super-$W_3^{(2)}$ algebra}

As subalgebras, $N=2$ SCA is linear, but
$W_3^{(2)}$ algebra is nonlinear. At the quantum level, $N=2$ SCA is the
same as the classical one, but $W_3^{(2)}$ algebra is different from the
classical one because of nonlinearity. Bershadsky has found quantum $W_3^{(
2)}$ algebra \cite{b} in the context of constrained Wess-Zumino-Witten model.
In order to extend the classical consideration \cite{ks} to the full quantum
version,
it is very natural to assume that
the {\it quantum} $W_3^{(2)}$ algebra \cite{b} and
$N=2$ SCA form
subalgebras in extended quantum $N=2$ super $W_3^{(2)}$ algebra.

As we could expect,
the algebraic structure of quantum
$W_3^{(2)}$ algebra\footnote{Normalizations of the currents are different from
those of ref. 2.}
is the same as the classical one except that $c$
dependent coefficients appearing in the right hand side of OPEs
are different\footnote{
For example, see the OPE of
$G^{+}(z_1)G^{-}(z_2)$.}. ( However, this is not the case for
$N=2$ super $W_3^{(2)}$ algebra, as we will see below. )
These two subalgebras take the form
\be
W_3^{(2)} : \left\{
\begin{array}{l}
J_w(z_1)J_w(z_2)  = \frac{1}{z_{12}^2}\frac{c}{6}  \quad , \quad
J_w(z_1)T_w(z_2)  =  \frac{1}{z_{12}^2}J_w,   \\
J_w(z_1)G^{+}(z_2)  =-  \frac{1}{z_{12}}\frac{1}{2}G^{+} \quad , \quad
T_w(z_1)G^{+}(z_2)  = \frac{1}{z_{12}^2}\frac{3}{2}G^{+}
+ \frac{1}{z_{12}}{G^{+}}', \\
T_w(z_1)T_w(z_2)=\frac{1}{z_{12}^4} \frac{(7-6c)c}{(3+2c)}+ \frac{1}{z_
{12}^2} 2 T_w+\frac{1}{z_{12}} T_w',  \\
 G^{+}(z_1)G^{-}(z_2)=\frac{1}{z_{12}^3} 2c-\frac{1}{z_{12}^2} 6 J_w+
\frac{1}{z_{12}}  \left[ -\frac{(3+2c)}{(-1+2c)} T_w+
\frac{24}{(-1+2c)} J_w J_w-
3 J_w'\right],
\end{array}
\right.
\ee
%1
\be
\mbox{N=2 SCA} : \left\{
\begin{array}{l}
J_s(z_1)J_s(z_2)  =  \frac{1}{z_{12}^2}\frac{c}{2} \quad , \quad
J_s(z_1)T_s(z_2)  =  \frac{1}{z_{12}^2}J_s,  \\
J_s(z_1)S(z_2)  =  \frac{1}{z_{12}}\frac{1}{2}S  \quad , \quad
 \\
S(z_1) \bar{S}(z_2)  =  \frac{1}{z_{12}^3}2c+\frac{1}{z_{12}^2}2J_s  +
        \frac{1}{z_{12}} \left[T_s+J_s'\right],   \\
T_s(z_1)S(z_2)  =  \frac{1}{z_{12}^2}\frac{3}{2}S+
                         \frac{1}{z_{12}}S' \quad , \quad
 \\
T_s(z_1)T_s(z_2)  =  \frac{1}{z_{12}^4}3c+\frac{1}{z_{12}^2}2T_s +
                         \frac{1}{z_{12}}T_s' .
\end{array} \right.
\ee
%2
where
\footnote{Hereafter we do not write down the regular OPEs.
All currents appearing in the right-hand sides of the OPEs are evaluated at
point $z_2$. Multiple composite currents are always regularized from the
right to the left, unless otherwise stated.
Also we have omitted the OPEs which can be obtained
through the following automorphism : $J_{w,s}\rightarrow -J_{w,s}$,
$G^{\pm}\rightarrow \pm G^{\mp}$, $S\rightarrow {\bar S}$,
${\bar S}\rightarrow S$, $S_1\rightarrow {\bar S}_1$,
${\bar S}_1\rightarrow -S_1$,$S_2\rightarrow -{\bar S}_2$,
${\bar S}_2\rightarrow  S_2$.}
$z_{12}=z_1-z_2$.
Let us also suppose that two $U(1)$
charges of currents for the quantum case
( with respect to the currents $J_s$ and $J_w$ ) are also the same as in the
classical case. Moreover, we assume that all the linear sub-algebras of the
classical case do {\it not} change their structure after passing to
the quantum case.
Therefore, they have the following form \cite{ks}:
\bea
& & S_1(z_1)\bar{S}_1(z_2)  =  -\frac{1}{z_{12}^2}\frac{c}{2}+
          \frac{1}{z_{12}} \frac{1}{2} \left[3J_w-J_s \right],
\quad
J_s(z_1)S_1(z_2)  =  -\frac{1}{z_{12}} \frac{1}{2}S_1, \nn \\
& & J_w(z_1)S_1(z_2)  =  -\frac{1}{z_{12}} \frac{1}{6} S_1, \quad
J_s(z_1)J_w(z_2)  =  \frac{1}{z_{12}^2} \frac{c}{3}, \quad
 J_s(z_1)T_w(z_2)  =  \frac{1}{z_{12}^2} 2J_w, \nn \\
& & J_s(z_1)G^{+}(z_2)  =  -\frac{1}{z_{12}} G^{+}, \quad
 J_s(z_1)S_2(z_2)  =  -\frac{1}{z_{12}} \frac{1}{2}S_2, \quad
J_w(z_1)T_s(z_2)  =  \frac{1}{z_{12}^2} \frac{2}{3}J_s, \nn \\
& &
J_w(z_1) S(z_2) = \frac{1}{z_{12}} \frac{1}{3} S, \quad
J_w(z_1)S_2(z_2)  =  -\frac{1}{z_{12}} \frac{1}{6}S_2, \quad
G^{-}(z_1)S_1(z_2)  = -
           \frac{1}{z_{12}} \frac{1}{2}S, \nn \\
& &
S_1(z_1)\bar{S}(z_2)  =
           \frac{1}{z_{12}} \frac{1}{2} G^{+}.
\eea
%3
It is natural to assume that all the remaining OPEs do not change their
structure except their structure constants
which we should fix from the Jacobi identities.
However, we have explicitly checked that the Jacobi identities
are not satisfied in this case.
In order to get the closed algebra for the quantum case
we should add extra terms
compared with classical one
\footnote{
We will explain later
that these extra terms disappear in the classical limit.}
to the right hand side of OPEs.
So we take the most general ansatz consistent with
the symmetry under the permutation
$z_1 \leftrightarrow z_2$,
statistics, spins, and conservations of two $U(1)$ charges.
As a result, we arrive at the following OPEs which satisfy
the Jacobi identities
for the generic value of the central charge
\bea
& & T_s(z_1)T_w(z_2)  =  \frac{1}{z_{12}^4} \frac{16c}{3(3+2c)}+\frac{1}{
z_{12}^2} \frac{1}{(3+2c)} \left[ 2 T_s-\frac{2(3+2c)}{(-1+2c)} T_w
+8 S_1\bar{S}_1+8 J_s J_w \right. \nn \\
& & \left. +\frac{48}{(-1+2c)} J_w J_w+2 J_s'-6 J_w' \right]
+\frac{1}{z_{12}} \frac{1}{(3+2c)} \left[ 4 S_1\bar{S}_2+4 S_1
\bar{S}_1'-4 S_2\bar{S}_1-\frac{6}{c} J_s J_w' \right. \nn \\
& & \left. +4 S_1'\bar{S}_1-\frac{2}{c} J_s' J_s+\frac{2(3+4c)}{c} J_s'
J_w+\frac{18}{c} J_w' J_w+2 T_s'+ J_s''-
3 J_w'' \right], \nn \\
& & T_s(z_1)G^{+}(z_2)  =  \frac{1}{z_{12}} \frac{2}{(-1+2c)}
  \left[ 2 S_1 \bar{S}-2 J_s G^{+}-{G^{+}}' \right], \nn \\
& & T_s(z_1)S_1(z_2)  =  -\frac{1}{z_{12}^2} \frac{1}{2} S_1+\frac{1}{z_{12}}
\frac{1}{2} \left[ S_2-\frac{1}{c} J_s S_1-\frac{3}{c} J_w S_1-S_1'
\right], \nn \\
& & T_s(z_1)S_2(z_2)  =  -\frac{1}{z_{12}^3} \frac{3(1+2c)}{2c} S_1+
\frac{1}{z_{12}^2} \left[ S_2+\frac{3}{2c} J_s S_1-\frac{9}{2c} J_w S_1-\frac{
3}{2} S_1' \right] \nn \\
& & +\frac{1}{z_{12}} \frac{1}{(-1+2c)} \left[ 2 G^{+} S-\frac{(1+6c)}{2c} J_s
S_2+\frac{(1+6c)}{2c^2} J_s J_s S_1+\frac{3(1+2c)}{c^2} J_s J_w S_1
\right. \nn \\
& &  +\frac{(-1+2c)}{c} J_s S_1'  -4 T_s S_1+\frac{3(-1+2c)}{2c} J_w
S_2-\frac{9(-1+2c)}{2c^2} J_w J_w
S_1 \nn \\
& & \left. -\frac{3(1+2c)}{c} J_w S_1'
+\frac{(-9+2c)}{2c} J_s' S_1
-\frac{3(3+2c)}{
2c} J_w' S_1+\frac{(3+2c)}{2} S_2'-\frac{(-1+2c)}{2} S_1'' \right], \nn \\
& & T_w(z_1)S_1(z_2)  =\frac{1}{z_{12}^2} \frac{(7-6c)}{6(3+2c)} S_1 \nn \\
& & +\frac{1}{z_{12}} \frac{1}{(3+2c)} \left[ \frac{(1-2c)}{2}
S_2+\frac{(-1+2c)}{2c}
J_s S_1-\frac{(3+10c)}{2c} J_w S_1+\frac{(1-2c)}{2} S_1' \right], \nn \\
& & T_w(z_1)S(z_2)  =
\frac{1}{z_{12}^2} \frac{8}{3(3+2c)} S +\frac{1}{z_{12}} \frac{4}{(3+2c)}
\left[
-G^{-} S_1+J_w S \right]  , \nn \\
& & T_w(z_1)S_2(z_2)  = \frac{1}{z_{12}^3} \frac{(-1+2c)(5+6c)}{2c(3+2c)}
S_1 \nn \\
& & +\frac{1}{z_{12}^2} \frac{1}{(3+2c)} \left[\frac{(11+6c)}{3} S_2-\frac{(
5+6c)}{2c} J_s S_1+\frac{3(5+6c)}{2c} J_w S_1
+\frac{(5+6c)}{2} S_1' \right] \nn \\
& & +\frac{1}{z_{12}} \frac{1}{(3+2c)} \left[ -2 G^{+} S-\frac{(-1+2c)}{2c}
J_s S_2
+\frac{(-1+2c)}{2c^2} J_s J_s S_1-\frac{3(1+2c)}{c^2} J_s J_w S_1
\right. \nn \\
& &-\frac{
(-1+2c)}{c} J_s S_1'
-\frac{4(3+2c)}{(-1+2c)} T_w S_1+\frac{(3+2c)}{2c} J_w S_2+\frac{3(3+12c+28
c^2)}{2c^2(-1+2c)} J_w J_w S_1 \nn \\
& &+\frac{3(1+2c)}{c} J_w S_1'-\frac{(-1+2c)}{2c}
J_s' S_1
+\frac{(-11+6c)(3+2c)}{2c(-1+2c)} J_w' S_1 \nn \\
& &\left. +\frac{(-1+2c)}{2} S_2'+
\frac{(-1+2c)}{2} S_1'' \right], \nn \\
& & G^{+}(z_1)S(z_2)  = -\frac{1}{z_{12}^2} 2 S_1+\frac{1}{z_{12}
} \left[ -S_2+\frac{1}{c} J_s S_1+\frac{3}{c} J_w S_1-S_1'\right], \nn \\
& & G^{+}(z_1)S_2(z_2)  =
           -\frac{1}{z_{12}} \frac{(5+6c)}{2(-1+2c)c}G^{+} S_1  , \nn \\
& & G^{-}(z_1)S_2(z_2)  = \frac{1}{z_{12}^2} \frac{(5+6c)}{4c} S \nn \\
& & +\frac{1}{z_{
12}} \left[ \frac{(5-2c)}{2c(-1+2c)} G^{-} S_1-\frac{1}{2c} J_s
S+\frac{3(1+6c)}{
2c(-1+2c)} J_w S+\frac{1}{2} S' \right], \nn \\
& & S_1(z_1)\bar{S}_2(z_2)  =-\frac{1}{z_{12}^3} \frac{1}{2}+\frac{1}{z_{12}
^2} \frac{1}{2c} \left[-J_s+3 J_w \right] \nn \\
& &+\frac{1}{z_{12}} \left[ \frac{1}{2} T_s+\frac{(3+2c)}{2(-1+2c)}
T_w+\frac{1}{
c} S_1 \bar{S}_1-\frac{1}{2c} J_s J_s-\frac{3(3+2c)}{2c(-1+2c)} J_w J_w
\right]
     , \nn \\
& & S(z_1)S_2(z_2)  =
           \frac{1}{z_{12}} \frac{3}{2c} S_1S, \nn \\
& & S(z_1)\bar{S}_2(z_2)  = -\frac{1}{z_{12}^2} \frac{3(1+2c)}{4c} G^{-}
\nn \\
& & +\frac{1}{z_{12}} \left[\frac{(3+2c)}{2c(-1+2c)} S \bar{S}_1
-\frac{(1+6c)}{2(-1+2c)c} J_s G^{-}+\frac{3}{2c} J_w G^{-}-\frac{1}{2} {G^{-}}'
\right] , \nn \\
& & S_2(z_1)S_2(z_2)  = \frac{1}{z_{12}} \frac{(1+2c)}{c(-1+2c)} \left[2 S_1
S_2-\frac{1}{c} S_1' S_1 \right]
            ,  \nn \\
& & S_2(z_1)\bar{S}_2(z_2)  = \frac{1}{z_{12}^4} \frac{(-1+4c+6c^2)}{2c}
+\frac{1}{z_{12}^3} \frac{(-1+4c+6c^2)}{2c^2} \left[ J_s-3 J_w \right] \nn \\
& & +\frac{1}{z_{12}^2} \frac{1}{(-1+2c)} \left[ \frac{(-1+3c+2c^2)}{c} T_s+
\frac{(3+2c)(-1+c-2c^2)}{c(-1+2c)} T_w+\frac{(-1+6c)}{c^2} S_1
\bar{S}_1 \right. \nn \\
& & +\frac{(-2+c)(-1+2c)}{2c^2} J_s J_s-\frac{3(1+6c)}{c} J_s J_w+
\frac{3(6-5c-4c^2+44c^2)}{2c^2(-1+2c)} J_w J_w \nn \\
& & \left. +\frac{(1+6c)}{2} J_s'-\frac{3(1+6c)}{2} J_w' \right]
 +\frac{1}{z_{12}}
\frac{1}{(-1+2c)} \left[2 G^{+} G^{-}+\frac{(-1+2c)}{c} S_1 \bar{S}_2+
2 S \bar{S} \right. \nn \\
& & +\frac{(-1+2c)}{c} S_2 \bar{S}_1+\frac{(-1+2c)}{c} J_s T_s
-\frac{(1+2c)(3+2c)}{c(-1+2c)} J_s T_w-\frac{(-1+2c)}{2c^2} J_s J_s J_s \nn \\
& &-\frac{3(-1+2c)}{2c^2} J_s J_s J_w+
\frac{3(3+4c+44c^2)}{2c^2(-1+2c)} J_s J_w
J_w-\frac{3(1+6c)}{2c} J_s J_w'
-\frac{3(1+2c)}{c} J_w T_s \nn \\
& &+\frac{3(3+2c)}{c} J_w T_w-\frac{9(3+10c)}{
2c^2} J_w J_w J_w+\frac{(1+2c)}{c^2} S_1' \bar{S}_1+\frac{(-1+2c)}{2c}
J_s' J_s \nn \\
& & -\frac{3(1+6c)}{2c} J_s' J_w
+\frac{3(-3+22c)}{2c} J_w' J_w+\frac{(-1+2c)}{2} T_s'-\frac{(3+2c)}{2} T_w' \nn
\\
& &\left. +
\frac{(-1+2c)}{2} J_s''-\frac{3(-1+2c)}{2} J_w'' \right]. \nn \\
\eea
%4
The full structure of our algebra
can be summarized as $(1),(2),(3),(4)$.
Several comments are in order here.
We now discuss the relationship between a classical algebra \cite{ks} and
 our quantized version. As we expected, the $c$-dependent structure
constants
 become more involved rational functions of $c$. Also one
can see that there exist {\it extra} terms in the right hand side of
$(4)$ which
do {\it not} appear in the
classical version.
The classical limit is given by the usual relation between the
Poisson bracket and the commutator while $c \rightarrow \infty $ \cite{bw}.
But we also need to take into account nonlinear terms because of
nonlinearity of our algebra. The straightforward way to recover the classical
limit is as follows.
If we consider any composite
current term
given by the product of $n$ fields in the right hand side of OPEs
at the classical level, then its
denominator should be proportional to the $(n-1)$th power of $c$.
So when we effect the $c \rightarrow \infty$ limit in any composite current,
product of $n$ fields, in quantum algebra, only the term that has the
above-mentioned property survives in the classical limit.
For example, look at the OPE of $T_s(z_1)T_w(z_2)$ in $(4
)$ and consider {\it only} the terms
$[2T_s/(3+2c)-2T_w/(-1+2c)]/z_{12}^2$
in the right hand side of it. It has "wrong" $c$-dependence and therefore
disappears in the classical limit.
Let us remark that all new terms compared to the classical case have
the "wrong" $c$-dependence and after applying these procedures to our
OPEs $(4)$, we recover the classical expressions \cite{ks}.

All the eight currents with spins less than two
are primary with respect to the following
Virasoro stress-tensor $T$ with zero central charge :
\be
T=\frac{1}{1+2c} \left[(-1+2c)T_s+(3+2c)T_w+8 S_1\bar{S}_1
-8 J_s^2+24 J_wJ_s-
  24 J_w^2+2 J_s'-6 J_w'\right],
\ee
%5
which also
reduces to the classical one \cite{ks} when $c \rightarrow \infty $.
$T_s$ and $T_w$ are the quasi primary fields with the central terms equal to
$3c$ and $(7-6c)c/(3+2c)$, respectively. However $S_2$ and $\bar{S_2}$ are
not ( quasi ) primary, they are primary in the following bases :
\bea
S_2 \rightarrow S_2+\frac{1}{2c} S_1', \quad
\bar{S_2} \rightarrow \bar{S_2}-\frac{1}{2c} \bar{S_1'} \nn \\
\eea
%6
It can be checked that in the quantum case also there is no
basis in our algebra such that all the currents
are primary with respect to  any spin two current satisfying
the Virasoro algebra.

Notice that the structure constants in the above algebra become divergent
if $c=0,1/2,$ or $-
3/2$.

{\bf 3. 'Hybrid' field realization }

Our analysis at the quantum level in this section is basically
the same as the one presented in \cite{ks}.
$N=2$ quantum super-$W_3^{(2)}$ algebra can be realized by the whole multiplets
of basic fields containing  six bosonic fields -
$\left\{ U_1,U_2,V_1,V_2,\xi,\bar\xi \right\}$ and six fermionic ones -
$\left\{ \lambda_1, {\bar\lambda}_1,\lambda_2, {\bar\lambda}_2,\psi,
\bar\psi \right\}$ with the spins $\left( 1,1,1,1,\frac{1}{2},\frac{1}{2}
\right)$, respectively and with the $J_s$- and $J_w$-charges equal to the
charges of corresponding currents.
The basic fields form the following superalgebra
\bea
& &\xi(z_1)\bar\xi(z_2) =  -\frac{1}{z_{12}} , \quad
\psi(z_1)\bar\psi(z_2)  =  -\frac{1}{z_{12}} , \quad
\lambda_1(z_1)\bar\lambda_1(z_2)  =  \frac{1}{z_{12}^2}+\frac{1}{z_{12}}
V_1, \nn \\
& &\lambda_2(z_1)\bar\lambda_2(z_2)  =
   \frac{1}{z_{12}^2}+\frac{1}{z_{12}}V_2, \quad
U_2(z_1)V_2(z_2)  =  -\frac{1}{z_{12}^2}, \quad
U_2(z_1)V_1(z_2)  =  -\frac{1}{z_{12}^2}, \nn \\
& &U_1(z_1)V_1(z_2)  =  \frac{1}{z_{12}^2}, \quad
U_1(z_1)\lambda_1(z_2)  =  \frac{1}{z_{12}}\lambda_1, \quad
U_1(z_1)\bar\lambda_1(z_2)  =  -\frac{1}{z_{12}}\bar\lambda_1, \quad \nn \\
& &U_2(z_1)\lambda_1(z_2)  =  -\frac{1}{z_{12}}\lambda_1, \quad
U_2(z_1)\bar\lambda_1(z_2)  =  \frac{1}{z_{12}}\bar\lambda_1, \quad
U_2(z_1)\lambda_2(z_2)  =  -\frac{1}{z_{12}}\lambda_2,\quad \nn \\
& &U_2(z_1)\bar\lambda_2(z_2)  =  \frac{1}{z_{12}}\bar\lambda_2. \nn \\
\eea
%7

Taking the most general ansatz for the
currents in terms of the defining basic fields and
demanding the consistency with the OPEs $(1),(2),(3),(4)$,  we can obtain the
following
realization of our algebra. We only write down the expressions for the
basic currents, $S,
{\bar S}, G^{+}, G^{-}$ because the remaining eight currents can
be obtained from the OPEs of basic ones.
\bea
S & = & 2^{1/2}\xi\bar\xi\psi - \bar\xi\bar
\lambda_2 +  \frac{(1 - 6c)}{2^{3/2}}V_1\psi
-  \frac{1}{2^{1/2}}V_2\psi +
  2^{1/2}U_1\psi +
  2^{1/2}U_2\psi +
  \frac{(-1 + 2c)}{2^{1/2}}\psi', \nn \\
{\bar S} & = & \frac{(1 + 2c)}{(-1 + 2c)^2}\xi\lambda_1 +
  \frac{2^{1/2}}{(1 - 2c)}\xi\bar\xi\bar\psi +
  \frac{1}{2^{1/2}(-1 + 2c)}V_1\bar\psi
+  \frac{(-3 + 2c)}{2^{3/2}(1 - 2c)}V_2\bar\psi \nn \\
         &   & +
  \frac{2^{1/2}}{(-1 + 2c)}U_1\bar\psi +
  \frac{1}{2^{1/2}}\bar\psi', \nn \\
G^{+} & = & \bar\lambda_1\bar\psi + \frac{2^{1/2}}{
(1 - 2c)}\xi\xi\bar\xi +
  \frac{2^{1/2}}{(1 - 2c)}\xi\psi\bar\psi
 +
  \frac{2^{1/2}(1+c)}{
   (-1 + 2c)}V_1\xi
+  \frac{(-3 + 2c)}{2^{3/2}(1 - 2c)}V_2\xi \nn \\
      &   & +
  \frac{2^{1/2}}{(-1 + 2c)}U_1\xi
+
  \frac{1}{2^{1/2}}\xi', \nn \\
G^{-} & = & (1+2c)  \lambda_2\psi + 2^{1/2}\xi
\bar\xi\bar\xi +
  2^{1/2}\bar\xi\psi\bar\psi
 +
  \frac{(1 - 6c)}{2^{3/2}}V_1\bar\xi
-  2^{1/2}(1 + c)V_2\bar\xi +
  2^{1/2}U_1\bar\xi  \nn \\
      &   & +
  2^{1/2}U_2\bar\xi +
  \frac{(-1 + 2c)}{2^{1/2}}\bar\xi',
\eea
%8
The above results (8) are defined up to possible automorphisms of
both the $N=2$ quantum $W_3^{(2)}$ algebra and the basic algebras (7).
It has been checked that we have correct classical limit \cite{ks} in
this 'hybrid' field realization through the following automorphism:
\bea
\xi=(2c)^{1/2} \tilde{\xi}, \quad \bar\xi=\frac{1}{(2c)^{1/2}} \tilde{\bar\xi},
\quad, \psi=\frac{1}{(2c)^{1/2}} \tilde{\psi}, \quad \bar\psi=(2c)^{1/2}
\tilde{
\bar\psi} \nn \\
\lambda_1=(2c)^{1/2} \tilde{\lambda_1}, \quad
\bar\lambda_1=\frac{1}{(2c)^{1/2}}
\tilde{\bar\lambda_1}, \quad \lambda_2=\frac{1}{(2c)^{1/2}} \tilde{\lambda_2},
\bar\lambda_2=(2c)^{1/2} \tilde{\bar\lambda_2}
\eea
%11
It is instructive to examine the structure of the
stress-tensor $T$ (5) in this realization.
It is bilinear in basic fields,
\bea
& &T=\left[-\lambda_1\bar\lambda_1 - \lambda_2\bar\lambda_2 -
\frac{1}{2}\xi\bar\xi' +
  \frac{1}{2}\psi\bar\psi' + \frac{1}{2}V_1V_1 + \frac{1}{2}V_2V_2 -
  V_2U_1 + U_1V_1 - U_2V_2 \right. \nn \\
& &\left. +
  \frac{1}{2}\xi'\bar\xi - \frac{1}{2}\psi'\bar\psi
 +
  \frac{(1 + 2c)}{4}V_1' + \frac{(3 - 2c)}{4}V_2' \right],
\eea
%12
and also reduces to its classical version as
$c\rightarrow\infty$.

{\bf 4. Conclusion}

To summarize, we have constructed the {\it quantum} $N=2$ super-$W_3^{(2)}$
algebra by using the Jacobi identities for which {\it extra} composite
currents in the right hand side of OPEs, that were {\it not} present in the
classical consideration, are crucial.
We have also presented its 'hybrid' field realization. The quantum $N=2$
super-$W_3^{(2)}$ has the same structure as the classical one: it
is a closure of quantum $N=2$ SCA and $W_3^{(2)}$ algebra. Thus, the
constructed $N=2$ super algebra contains $W_3^{(2)}$
as a genuine subalgebra in contrast to the known $N=2$
superextension of $W_3$ algebra which yields $W_3$ only in the limit of
vanishing fermionic currents.

Despite the presence of $N=2$ SCA as subalgebra and the equal numbers of
bosonic and fermionic currents in $N=2$ quantum super-$W_3^{(2)}$ algebra,
the spin contents of currents and OPEs make it impossible to combine
{\it ad hoc}
the currents into $N=2$ supermultiplets (the currents of
$N=2$ SCA appear in the right hand side of OPEs between other currents).
It would be very interesting to study manifestly $N=2$ supersymmetric
formulation of this algebra which allows us to
combine the currents into $N=2$ supermultiplets. The main idea of such
reformulation is to look for another $N=2$ SCA in the full $N=2$
super-$W_3^{(2)}$ algebra. In the forthcoming papers \cite{sf1,sf2}
we will present
the corresponding $N=2$ superfield formulations of $N=2$ $W_3^{(2)}$
algebra both at the classical and quantum levels.

{\bf Acknowledgments}

We would like to thank  S.Bellucci, A.Pashnev and
especially E.Ivanov for many
useful and clarifying discussions and V.Ogievetsky for his interest in this
work.

\end{document}